\documentclass[twocolumn,prb,aps,showpacs,amsmath,amssymb]{revtex4}

\usepackage{graphicx}
\usepackage{dcolumn}
\usepackage{bm}

\renewcommand{\vec}[1]{{\bf#1}}

\begin{document}

\title{Charge ratchet from spin flip: space-time symmetry paradox}

\author{Sergey Smirnov,$^1$ Dario Bercioux,$^2$ Milena Grifoni,$^1$ and Klaus Richter$^1$}
\affiliation{$^1$Institut f\"ur Theoretische Physik, Universit\"at Regensburg, D-93040 Regensburg, Germany\\
  $^2$Freiburg Institute for Advanced Studies (FRIAS) and Physikalisches Institut, Universit\"at Freiburg, D-79104
  Freiburg, Germany}

\date{\today}

\begin{abstract}
Traditionally the charge ratchet effect is considered as a consequence of either the spatial symmetry breaking engineered
by asymmetric periodic potentials, or time asymmetry of the driving fields. Here we demonstrate that electrically and
magnetically driven quantum dissipative systems with spin-orbit interactions represent an exception from this standard
idea. In contrast to the so far well established belief, a charge ratchet effect appears when both the periodic potential
and driving are symmetric. We show that the source of this paradoxical charge ratchet mechanism is the coexistence of
quantum dissipation with the spin flip processes induced by spin-orbit interactions.
\end{abstract}


\maketitle

\section{Introduction}
A system of particles in a periodic potential and driven by a time-dependent external force may exhibit a net current even
if the force has zero time average. This so-called particle ratchet effect
\cite{Astumian,Reimann,Juelicher,Linke,Majer,Haenggi,Olbrich} is used e.g. in nano-generators of direct charge currents
\cite{Linke,Olbrich}. To excite the particle ratchet current it is traditionally believed that the asymmetry of either the
periodic potential or driving force is a must. In the quantum regime a more stringent conclusion has been obtained: in
quantum systems in which charged particles populate only one Bloch band the charge ratchet effect does not exist, even if
the periodic potential is asymmetric, unless time asymmetry is provided by the driving field \cite{Grifoni,Goychuk}.
Indeed, the ratchet effect exists in a single-band system which is driven by a field with harmonic mixing
\cite{Goychuk,Ponomarev}.

The concepts and conclusions mentioned above are based on considering particles as spinless, that is without taking into
account any possible impact from switching between the spin states of the particles involved in ratchet transport. In
various physical systems there is a plenty of ways to change the spin states of a particle. In this paper we limit
ourselves to semiconductor heterostructures with spin-orbit interactions since from the practical point of view these
systems are attractive for fabrication of nano-devices.

For semiconductor heterostructures with spin-orbit interactions, described for example by Rashba \cite{Rashba} or
Dresselhaus \cite{Dresselhaus} spin-orbit Hamiltonians, the {\it spin} ratchet effect is rooted in an asymmetric
excitation of spin dynamics by the orbital dynamics induced by an electric field. For electrically driven coherent and
dissipative systems with Rashba spin-orbit interaction (RSOI) the spin ratchet mechanism has been confirmed
\cite{Scheid,Smirnov,Smirnov_1}. Even for symmetric periodic potentials and symmetric driving the spin ratchet effect
exists \cite{Scheid}. However, the {\it charge} ratchet effect is absent in both the coherent and dissipative cases when
both the periodic potential and driving force are symmetric. This could deepen the impression that a system with symmetric
periodic potentials will never respond to time-symmetric external fields via the charge ratchet mechanism and systems with
spin-orbit interactions like all other systems obey this habitual rule. The present work reveals that this is a delusion
and in reality systems with spin-orbit interactions provide a unique opportunity to answer the fundamental questions
related to the role of symmetries in the charge ratchet phenomena in general.

In this paper we show that the space asymmetry of the periodic potentials and the time asymmetry of the driving fields,
usually required as key properties of charge ratchets, are not necessary as the Rashba spin flip processes alone are
sufficient even if a dissipative system is time-symmetrically driven. Specifically, it is found that the charge ratchet
effect in this case exists for space-symmetric periodic potentials and time-symmetric driving by electric {\it and}
magnetic fields. It stems just from the simultaneous presence of quantum dissipation and the spin flip processes of Rashba
electrons. The ratchet charge current in the system is unusual. Its queerness consists in the fact that this current, in
contrast to early predictions for systems without spin-orbit interactions \cite{Grifoni,Goychuk}, appears even when only
one energy band provides electrons for transport and no harmonic mixing is present in the driving fields. This charge
current is of pure spin-orbit nature and, as a result, it disappears when the spin-orbit coupling strength vanishes.
Therefore such spin-orbit charge currents can be controlled by the same gate voltage which controls the strength of the
spin-orbit coupling in the system. It is evident that this peculiarity of the charge ratchet current is very attractive
from the experimental point of view.

The paper is organized as follows. Section \ref{Model} presents the model which is solved in Section \ref{Solution} and
numerically analyzed in Section \ref{NR}. Section \ref{Conclusion} concludes the paper.
\section{Model}\label{Model}
\begin{figure}
\includegraphics[width=8.5 cm]{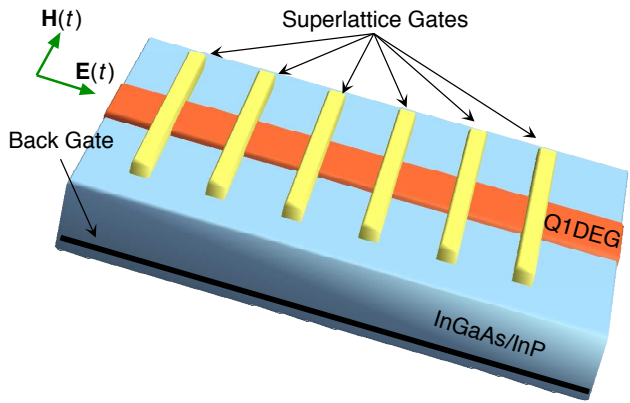}
\caption{\label{figure_1} (Color online) A 2DEG with RSOI of strength $\alpha=9.94\times 10^{-12}$ eV$\cdot$m is
obtained by a gate voltage applied to an InGaAs/InP heterostructure using the "Back Gate". The electron effective mass is
$m=0.037m_0$ with $m_0$ being the free electron mass and the effective gyroscopic factor is $g^*=-15$. A parabolic
confinement of strength $\hbar\omega_0=0.225$ meV forms in the 2DEG a quasi-one-dimensional electron gas (Q1DEG). The
superlattice with period $L=0.25\,\mu m$ is shaped by the "Superlattice Gates" which create a symmetric periodic
potential. The system is driven by a longitudinal electric field $\vec{E}(t)$ and by a transverse magnetic field
$\vec{H}(t)$ which are time-symmetric.}
\end{figure}

An archetype of the device under investigation is shown in Fig.~\ref{figure_1}. In this system non-interacting electrons
are confined in a quasi-one-dimensional (quasi-1D) periodic structure obtained by appropriately placed gates applied to a
two-dimensional electron gas (2DEG) with RSOI. The system interacts with an external environment (or bath): the
longitudinal orbital degree of freedom of each electron is coupled to orbital degrees of freedom of the external
environment. This coupling is the source of dissipation in the system. The electrons are driven by longitudinal electric
and transverse in-plane magnetic homogeneous fields which are time-symmetric and time-periodic functions with zero mean
value.

To perform a quantitative analysis of the charge ratchet effect we model the system by the Hamiltonian
$\hat{H}(t)=\hat{H}_0+\hat{H}_\mathrm{D}(t)+\hat{H}_\mathrm{B}$, where $\hat{H}_\mathrm{D}(t)\equiv -eE(t)\hat{x}-
g\mu_\mathrm{B}H(t)\hat{\sigma}_z$ is the driving term, $\hat{H}_\mathrm{B}$ is the bath term of the Caldeira-Leggett model
\cite{Caldeira,Weiss} taking into account the orbital coupling between the electron longitudinal degree of freedom,
$\hat{x}$, and orbital degrees of freedom of the bath. All properties of the bath are encapsulated in its spectral density
$J(\omega)$. Finally, $\hat{H}_0$ is the Hamiltonian of the isolated system:
\begin{equation}
\hat{H}_0\equiv\frac{\hbar^2\hat{\vec{k}}^2}{2m}-
\frac{\hbar^2k_{\mathrm{so}}}{m}\bigl(\hat{\sigma}_x\hat{k}_z-\hat{\sigma}_z\hat{k}_x\bigl)+V(\hat{x},\hat{z}),
\label{HIS}
\end{equation}
where $V(x,z)\equiv m\omega_0^2z^2/2+U(x)$ and $U(x)=U(-x)$. In this model it is assumed that the 2DEG is in the $x-z$
plane and the quasi-1D structure is formed along the $x$-axis using a harmonic confinement of strength $\omega_0$ along the
$z$-axis. The electron spin $g$-factor is denoted as $g$ and $\mu_\mathrm{B}$ is the Bohr magneton. The super-lattice
period is $L$, $U(x+L)=U(x)$. The parameter $k_\mathrm{so}\equiv\alpha m/\hbar^2$ characterizes the strength of the
spin-orbit coupling.

The electric driving is given by the vector $\vec{E}(t)=(E(t),0,0)$ while the magnetic driving is $\vec{H}(t)=(0,0,H(t))$.
We consider the symmetric time dependence: $eE(t)\equiv F\cos(\Omega(t))$, $H(t)\equiv H\cos(\Omega(t))$. The vector
potential is chosen using the Landau gauge $\vec{A}(t)=(-H(t)y,0,0)$. Since $y=0$ in the 2DEG, the vector potential is not
explicitly present in the model.

We would like to mention that the in-plane electric fields corresponding to $U(x)$, $m\omega_0^2z^2/2$ and the driving
electric field $E(t)$ are assumed to be much weaker than the out-of plane electric field forming the 2DEG with RSOI. Thus
they produce very weak, in comparison with RSOI, spin-orbit interactions which, therefore, may be neglected.
\section{Solution}\label{Solution}
Before starting a rigorous exploration one can already anticipate that the magnetic field driving brings a whiff of fresh
physics because the spin dynamics can be controlled directly and not only through the spin-orbit interaction mediating
between the electric field and electron spins.

To study the charge ratchet effect at low temperatures, when only the lowest Bloch band of the super-lattice is populated
with electrons, we calculate the charge current averaged over one driving period. This current in the long time limit
provides the stationary charge ratchet response of the system. The common eigenstates of $\hat{x}$ and $\hat{\sigma}_z$
represent a convenient basis to obtain this response. Because of the discrete eigenvalue structure of $\hat{x}$ (see
below) the basis is called the $\sigma$-discrete variable representation ($\sigma$-DVR) basis. The eigenstates are denoted
as $|m,j,\sigma\rangle$, where $m=0,\pm 1,\pm 2,\ldots$, and $j$ and $\sigma$ are the transverse mode and spin quantum
numbers, respectively \cite{Smirnov,Smirnov_1}. Below, in parallel with our main goal for this paper, that is the charge
ratchet current, we also provide the results for the spin ratchet current to show that, as in the coherent case
\cite{Scheid}, it also exists in a dissipative system with symmetric periodic potentials and symmetric driving. In the
$\sigma$-DVR basis the averaged charge and spin currents have a simple form \cite{Smirnov,Smirnov_1}:
\begin{equation}
\begin{split}
&J_\mathrm{C}=-e\underset{t\to\infty}{\mathrm{lim}}\sum_{m,j,\sigma} x_{m,j}\frac{d}{dt}P_{j,\sigma}^m(t),\\
&J_\mathrm{S}=\underset{t\to\infty}{\mathrm{lim}}\sum_{m,j,\sigma}\sigma x_{m,j}\frac{d}{dt}P_{j,\sigma}^m(t).
\end{split}
\label{CCSC}
\end{equation}
In Eq. (\ref{CCSC}) $P_{j,\sigma}^m(t)$ is the averaged population at time $t$ of the $\sigma$-DVR state
$|m,j,\sigma\rangle$, the quantities $x_{m,j}=mL+d_j$ ($-L/2<d_j\leqslant L/2$) and $\sigma$ are eigenvalues of $\hat{x}$
and $\hat{\sigma}_z$ corresponding to their common eigenstate $|m,j,\sigma\rangle$. Note that in Eq. (\ref{CCSC}) one has
to first calculate the sum and only afterwards to take the limit because the operations of taking limit and infinite
summation do not commute as it was proven in Refs. \onlinecite{Smirnov,Smirnov_1}. Additionally, the $\sigma$-DVR basis
allows the path integral formalism to handle the magnetic driving on an equal footing with the standard electric driving
since in this basis the whole driving Hamiltonian, $\hat{H}_\mathrm{D}(t)$, is diagonal.

In the long time limit the populations $P_{j,\sigma}^m(t)$ come from a master equation \cite{Smirnov,Weiss} which is in this
case Markovian.

An analytical treatment of this rather complicated problem is possible when the dynamics of $P_{j,\sigma}^m(t)$ is treated
within the first two transverse modes, i.e., $j=0,1$.

For a detailed study we derive the charge and spin currents assuming that the hopping matrix elements between neighboring
$\sigma$-DVR states are small. Following the steps thoroughly described in Ref. \onlinecite{Smirnov_1} we obtain:
\begin{equation}
\begin{split}
&J_\mathrm{C}\!=\!
\frac{2eL}{I}\bigl|\Delta_{\uparrow\downarrow}^{01}\bigl|^2\bigl|\Delta_{\downarrow\uparrow}^{10}\bigl|^2
\bigl(I_{\uparrow\downarrow}^{01,\mathrm{b}}I_{\downarrow\uparrow}^{10,\mathrm{b}}-
I_{\uparrow\downarrow}^{01,\mathrm{f}}I_{\downarrow\uparrow}^{10,\mathrm{f}}\bigl),\\
&J_\mathrm{S}\!=\!
\frac{2L}{I}\bigl(\bigl|\Delta_{\uparrow\downarrow}^{01}\bigl|^4I_{\uparrow\downarrow}^{01,\mathrm{f}}
I_{\downarrow\uparrow}^{10,\mathrm{b}}-\bigl|\Delta_{\downarrow\uparrow}^{10}\bigl|^4I_{\uparrow\downarrow}^{01,\mathrm{b}}
I_{\downarrow\uparrow}^{10,\mathrm{f}}\bigl),
\end{split}
\label{CCSCF}
\end{equation}
where $\Delta_{\sigma'\sigma}^{j'j}\equiv\,\langle m+1,j',\sigma'|\hat{H}_0|m,j,\sigma\rangle$ are the hopping
matrix elements of the Hamiltonian of the isolated system, Eq. (\ref{HIS}),
$I\!\equiv\!\bigl|\Delta_{\uparrow\downarrow}^{01}\bigl|^2\bigl(I_{\uparrow\downarrow}^{01,f}\!\!+
\!I_{\downarrow\uparrow}^{10,b}\bigl)+\bigl|\Delta_{\downarrow\uparrow}^{10}\bigl|^2\bigl(I_{\uparrow\downarrow}^{01,b}\!+
\!I_{\downarrow\uparrow}^{10,f}\bigl)$, and $\uparrow,\downarrow$ stand for $\sigma=1,-1$, respectively. The effects of both
the driving fields and quantum dissipation are in the integrals \cite{Weiss}
\begin{equation}
\begin{split}
I_{\sigma'\sigma}^{j'j,\mathrm{\binom{f}{b}}}
&\equiv\frac{1}{\hbar^2}\int_{-\infty}^{\infty}d\tau\mathrm{e}^{-\frac{L^2}{\hbar}
Q(\tau;J(\omega),T)+\mathrm{i}\frac{\tau}{\hbar}(\varepsilon_\sigma^j-\varepsilon_{\sigma'}^{j'})}\times\\
&\times J_0\biggl[\frac{\mp 2FL+2g\mu_\mathrm{B}H(\sigma-\sigma')}{\hbar\Omega}
\sin\biggl(\frac{\Omega\tau}{2}\biggl)\biggl],
\end{split}
\label{DDI}
\end{equation}
where $Q[\tau;J(\omega),T]$ is the twice integrated bath correlation function,
\begin{equation}
\begin{split}
Q(\tau)\equiv\frac{1}{\pi}\int_0^\infty\,&d\omega\frac{J(\omega)}{\omega^2}
\biggl[\coth\biggl(\frac{\hbar\omega}{2k_\mathrm{Boltz.}T}\biggl)\times\\
&\times[1-\cos(\omega\tau)]+\mathrm{i}\sin(\omega\tau)\biggl],
\end{split}
\label{TIBCF}
\end{equation}
whose dependence on $\tau$ is fixed by the bath spectral density $J(\omega)$ and temperature $T$,
$\varepsilon_\sigma^j\equiv\langle m,j,\sigma|\hat{H}_0|m,j,\sigma\rangle$ are the on-site energies of the isolated system,
and $J_0(x)$ is the Bessel function of zero order.

Remarkably, Eq. (\ref{CCSCF}) tells us that at low temperatures the ratchet charge and spin transport in the system exists
just because of spin flip processes. Whereas it looks natural for the spin current, it is a quite unexpected and important
result for the charge current. This current emerges because the magnetic driving changes the charge dynamics. In this case
the spin-orbit interaction plays a role inverse to the one which it plays for the electric driving: the magnetic field
exciting spin dynamics induces orbital dynamics through the spin-orbit interaction. The corresponding charge flow,
originating just due to the spin-orbit interaction, is finite even when only one Bloch band contributes to transport.

\begin{figure}
\includegraphics[width=7.2 cm]{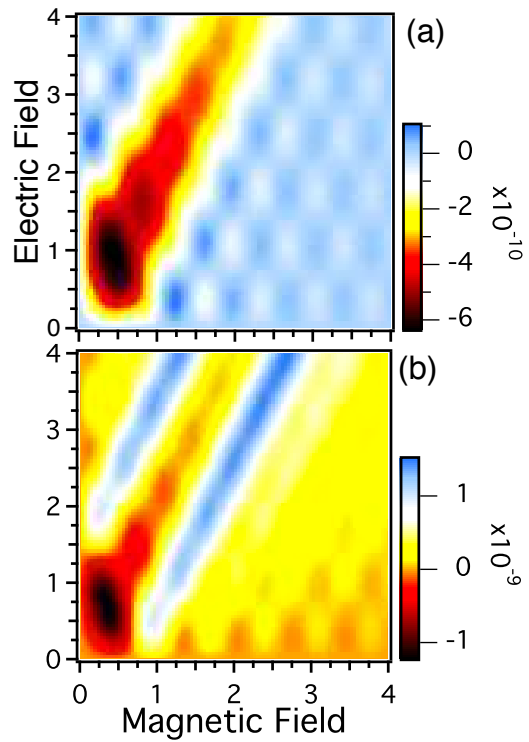}
\caption{\label{figure_2} (Color online) The charge and spin ratchet currents as functions of the amplitudes of the
  electric and magnetic fields. a, Charge current. b, Spin current. The amplitudes of the electric, $FL$, and magnetic,
  $g\mu_\mathrm{B}H$, fields are in units of $\hbar\omega_0$. The currents are in units of $L\omega_0$. According to
  Eq. (\ref{RC}) the charge and spin currents are excited when both the electric and magnetic fields simultaneously drive
  the system.}
\end{figure}
At this point it is important to note that since electrons populate only one Bloch band, the spatial asymmetry is not
enough when the driving is time-symmetric. This is in complete accordance with the results obtained earlier
\cite{Grifoni,Goychuk} for spinless particles and is clearly demonstrated in our case by the structure of the Rashba
Hamiltonian. Indeed, this Hamiltonian has two terms, $\hat{\sigma}_z\hat{k}_x$ and $\hat{\sigma}_x\hat{k}_z$. The first
term does not flip the electron spin and does not lead to the charge ratchet effect while the second one flips the
electron spin. It is exactly this second term which is responsible for the paradoxical situation: charge ratchet effect
for a space-symmetric periodic potential, time-symmetric driving and one Bloch band transport. The charge ratchet effect
is exclusively based on the spin-flip processes in the isolated system and thus it is fundamentally different from the
charge ratchet mechanisms which have been known so far.

The situation, however, is highly non-trivial and the final conclusions about the existence of the ratchet charge and spin
flows cannot be based only on the presence of spin-orbit interactions. There are also external time-dependent fields
driving the system and internal quantum dissipative processes. The mutual driving-dissipation effect is incorporated in
the integrals, Eq. (\ref{DDI}). Therefore, a further analysis is required: one should additionally take into consideration
the properties of the integrals from Eq. (\ref{DDI}) and the properties of the static periodic potential with respect to
the spatial inversion symmetry.

This analysis leads to the following results:
\begin{equation}
\begin{split}
&F\neq 0,\,H=0\quad\Longrightarrow\quad J_\mathrm{C}=0,\,J_\mathrm{S}=0,\\
&F=0,\,H\neq 0\quad\Longrightarrow\quad J_\mathrm{C}=0,\,J_\mathrm{S}=0,\\
&F\neq 0,\,H\neq 0\quad\Longrightarrow\quad J_\mathrm{C}\neq 0,\,J_\mathrm{S}\neq 0.
\end{split}
\label{RC}
\end{equation}

The results presented in Eq. (\ref{RC}) are easily obtained from Eq. (\ref{CCSCF}) if one takes into account that for
$U(x)=U(-x)$ the equality $|\Delta_{\uparrow\downarrow}^{01}\bigl|=|\Delta_{\downarrow\uparrow}^{10}\bigl|$ is valid
\cite{Smirnov,Smirnov_1}, and for $F=0$ or $H=0$ one makes use of the equality
$I_{\sigma'\sigma}^{j'j,\mathrm{f}}=I_{\sigma',\sigma}^{j'j,\mathrm{b}}$ which follows from Eq. (\ref{DDI}).

The principal feature of the physics taking place when $F\neq 0$ and $H\neq 0$ is that the existence of the ratchet
effects is {\it not} dictated only by properties of the isolated system as in Refs. \onlinecite{Smirnov,Smirnov_1}. The
physical picture is now more intricate. In the charge and spin currents one cannot find clear traces of either driving and
dissipation or the isolated system. The two imprints are not separable and the charge and spin ratchet mechanisms are
determined by the whole system-plus-bath complex. Note that in comparison with the spin ratchet current in Refs.
\onlinecite{Smirnov,Smirnov_1} the charge ratchet current in Eq. (\ref{CCSCF}) factorizes into two factors in a different
way. While in the spin ratchet current in Refs. \onlinecite{Smirnov,Smirnov_1} there was a factor representing a difference
of the hopping matrix elements of the Hamiltonian of the isolated system, now in the charge ratchet current there is a
factor representing the difference $I_{\uparrow\downarrow}^{01,\mathrm{b}}I_{\downarrow\uparrow}^{10,\mathrm{b}}-
I_{\uparrow\downarrow}^{01,\mathrm{f}}I_{\downarrow\uparrow}^{10,\mathrm{f}}$ which is not related only to the isolated system. As
one can see from Eq. (4), this difference takes into account the combined effect of dissipation through the twice
integrated bath correlation function, driving through the Bessel function and isolated system through the on-site energies
storing information about the periodic potential. In the same way as the difference of the hopping matrix elements of the
Hamiltonian of the isolated system in Refs. \onlinecite{Smirnov,Smirnov_1} dictated the existence of the spin ratchet
current, now the difference $I_{\uparrow\downarrow}^{01,\mathrm{b}}I_{\downarrow\uparrow}^{10,\mathrm{b}}-
I_{\uparrow\downarrow}^{01,\mathrm{f}}I_{\downarrow\uparrow}^{10,\mathrm{f}}$ dictates the existence of the charge ratchet current
in the present paper and results in the combined effect of the isolated system, dissipation and driving, as mentioned
above. It is important to remember that this combined effect takes place only if the spin-orbit coupling is finite because
$\Delta_{\uparrow\downarrow}^{01}=\Delta_{\downarrow\uparrow}^{10}=0$ in the absence of RSOI, as it has been proven in Refs.
\onlinecite{Smirnov,Smirnov_1}.

\begin{figure}
\includegraphics[width=8.5 cm]{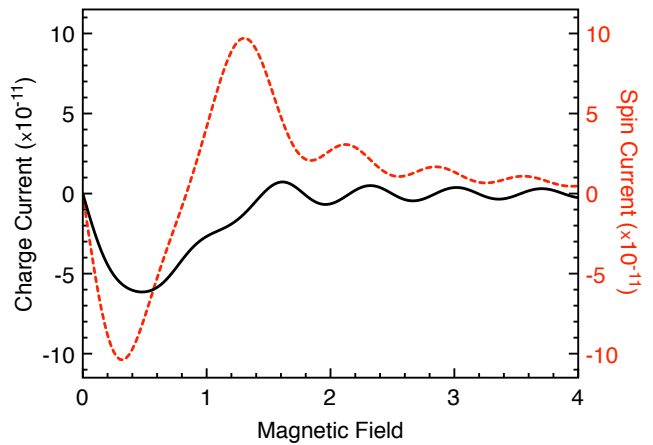}
\caption{\label{figure_3} (Color online) The charge (solid curve) and spin (dashed curve) ratchet currents as functions of
  the magnetic field amplitude. The magnetic amplitude, $g\mu_\mathrm{B}H$, is in units of $\hbar\omega_0$. The electric
  amplitude is fixed, $FL=\hbar\omega_0$. The currents oscillate and have non-universal zero points which depend on
  concrete values of the physical parameters of the system-plus-bath complex.}
\end{figure}
\section{Numerical results}\label{NR}
Numerical calculations based on Eqs. (\ref{CCSCF}) and (\ref{DDI}) have been performed to obtain the dependence of the
ratchet currents on $F$ and $H$. Figure \ref{figure_2} shows this dependence. The superlattice is modeled by the symmetric
potential $U(x)=2.6\hbar\omega_0[1-\cos(2\pi x/L)]$. The period is $L=2.5\sqrt{\hbar/m\omega_0}$ which gives
$k_\mathrm{so}L\approx 0.368\pi$. The driving frequency of the electric and magnetic fields is $\Omega=\sqrt{3}\omega_0/4$.
The bath is Ohmic with the exponential cut-off at $\omega_c=10\omega_0$: $J(\omega)=\eta\omega\exp(-\omega/\omega_c)$. The
viscosity coefficient is $\eta=0.1m\omega_0$ and the temperature is $k_\mathrm{B}T=0.5\hbar\omega_0$. As expected from Eq.
(\ref{RC}) the ratchet effects exist for the space-symmetric periodic potential and time-symmetric driving. From
Fig.~\ref{figure_2} one also observes an oscillatory behavior of the ratchet currents.

These oscillations are detailed in Fig.~\ref{figure_3}. As one can see the currents can be equal to zero even when both of
the driving fields are finite. These zero-current points are not universal: they depend on concrete values of the
physical parameters of the isolated system and bath. In contrast, the conditions in Eq. (\ref{RC}) are universal, i.e.,
they do not depend on concrete values of the physical parameters of the semiconductor heterostructure and environment.

Finally, we would like to note that since our theory is a theory of a strongly dissipative tight-binding system, the charge
ratchet current is small but detectable. For example using the parameters of Ref. \onlinecite{Smirnov_1} we get the charge
ratchet current $J_\mathrm{C}\sim 10$ fA. We expect that models with weak dissipation or/and weak periodic potentials will
give much larger charge ratchet currents in the fully symmetric setup presented in this paper.
\section{Conclusion}\label{Conclusion}
In summary, in contrast to the common belief, we have shown that the existence of spin flip processes in a dissipative
system is already sufficient to produce the charge ratchet effect even if the periodic potential is space-symmetric and
the system is driven by time-symmetric fields. To be specific we have considered Rashba spin-orbit interaction as a
mechanism for the electron spin flip. The charge ratchet current has been found to have a purely spin flip origin. The
space asymmetry of the periodic potential and the time asymmetry of the driving fields have not been necessary.

\begin{acknowledgments}
Support from the DFG under the program SFB 689 and Excellence Initiative of the German Federal and State Governments is
acknowledged.
\end{acknowledgments}


\begin{thebibliography}{17}
\expandafter\ifx\csname natexlab\endcsname\relax\def\natexlab#1{#1}\fi
\expandafter\ifx\csname bibnamefont\endcsname\relax
  \def\bibnamefont#1{#1}\fi
\expandafter\ifx\csname bibfnamefont\endcsname\relax
  \def\bibfnamefont#1{#1}\fi
\expandafter\ifx\csname citenamefont\endcsname\relax
  \def\citenamefont#1{#1}\fi
\expandafter\ifx\csname url\endcsname\relax
  \def\url#1{\texttt{#1}}\fi
\expandafter\ifx\csname urlprefix\endcsname\relax\def\urlprefix{URL }\fi
\providecommand{\bibinfo}[2]{#2}
\providecommand{\eprint}[2][]{\url{#2}}

\bibitem[{\citenamefont{Astumian and H{\"a}nggi}(2002)}]{Astumian}
\bibinfo{author}{\bibfnamefont{R.~D.} \bibnamefont{Astumian}} \bibnamefont{and}
  \bibinfo{author}{\bibfnamefont{P.}~\bibnamefont{H{\"a}nggi}},
  \bibinfo{journal}{Phys.\ Today} \textbf{\bibinfo{volume}{55}},
  \bibinfo{pages}{33} (\bibinfo{year}{2002}).

\bibitem[{\citenamefont{Reimann et~al.}(1997)\citenamefont{Reimann, Grifoni,
  and H{\"a}nggi}}]{Reimann}
\bibinfo{author}{\bibfnamefont{P.}~\bibnamefont{Reimann}},
  \bibinfo{author}{\bibfnamefont{M.}~\bibnamefont{Grifoni}}, \bibnamefont{and}
  \bibinfo{author}{\bibfnamefont{P.}~\bibnamefont{H{\"a}nggi}},
  \bibinfo{journal}{Phys.\ Rev.\ Lett.} \textbf{\bibinfo{volume}{79}},
  \bibinfo{pages}{10} (\bibinfo{year}{1997}).

\bibitem[{\citenamefont{J{\"u}licher et~al.}(1997)\citenamefont{J{\"u}licher,
  Ajdari, and Prost}}]{Juelicher}
\bibinfo{author}{\bibfnamefont{F.}~\bibnamefont{J{\"u}licher}},
  \bibinfo{author}{\bibfnamefont{A.}~\bibnamefont{Ajdari}}, \bibnamefont{and}
  \bibinfo{author}{\bibfnamefont{J.}~\bibnamefont{Prost}},
  \bibinfo{journal}{Rev.\ Mod.\ Phys.} \textbf{\bibinfo{volume}{69}},
  \bibinfo{pages}{1269} (\bibinfo{year}{1997}).

\bibitem[{\citenamefont{Linke et~al.}(1999)\citenamefont{Linke, Humphrey,
  L{\"o}fgren, Sushkov, Newbury, Taylor, and Omling}}]{Linke}
\bibinfo{author}{\bibfnamefont{H.}~\bibnamefont{Linke}},
  \bibinfo{author}{\bibfnamefont{T.~E.} \bibnamefont{Humphrey}},
  \bibinfo{author}{\bibfnamefont{A.}~\bibnamefont{L{\"o}fgren}},
  \bibinfo{author}{\bibfnamefont{A.~O.} \bibnamefont{Sushkov}},
  \bibinfo{author}{\bibfnamefont{R.}~\bibnamefont{Newbury}},
  \bibinfo{author}{\bibfnamefont{R.~P.} \bibnamefont{Taylor}},
  \bibnamefont{and} \bibinfo{author}{\bibfnamefont{P.}~\bibnamefont{Omling}},
  \bibinfo{journal}{Science} \textbf{\bibinfo{volume}{286}},
  \bibinfo{pages}{2314} (\bibinfo{year}{1999}).

\bibitem[{\citenamefont{Majer et~al.}(2003)\citenamefont{Majer, Peguiron,
  Grifoni, Tusveld, and Mooij}}]{Majer}
\bibinfo{author}{\bibfnamefont{J.~B.} \bibnamefont{Majer}},
  \bibinfo{author}{\bibfnamefont{J.}~\bibnamefont{Peguiron}},
  \bibinfo{author}{\bibfnamefont{M.}~\bibnamefont{Grifoni}},
  \bibinfo{author}{\bibfnamefont{M.}~\bibnamefont{Tusveld}}, \bibnamefont{and}
  \bibinfo{author}{\bibfnamefont{J.~E.} \bibnamefont{Mooij}},
  \bibinfo{journal}{Phys.\ Rev.\ Lett.} \textbf{\bibinfo{volume}{90}},
  \bibinfo{pages}{056802} (\bibinfo{year}{2003}).

\bibitem[{\citenamefont{H{\"a}nggi and Marchesoni}(2009)}]{Haenggi}
\bibinfo{author}{\bibfnamefont{P.}~\bibnamefont{H{\"a}nggi}} \bibnamefont{and}
  \bibinfo{author}{\bibfnamefont{F.}~\bibnamefont{Marchesoni}},
  \bibinfo{journal}{Rev.\ Mod.\ Phys.} \textbf{\bibinfo{volume}{81}},
  \bibinfo{pages}{387} (\bibinfo{year}{2009}).

\bibitem[{\citenamefont{Olbrich et~al.}(2009)\citenamefont{Olbrich, Ivchenko,
  Ravash, Feil, Danilov, Allerdings, Weiss, Schuh, Wegscheider, and
  Ganichev}}]{Olbrich}
\bibinfo{author}{\bibfnamefont{P.}~\bibnamefont{Olbrich}},
  \bibinfo{author}{\bibfnamefont{E.~L.} \bibnamefont{Ivchenko}},
  \bibinfo{author}{\bibfnamefont{R.}~\bibnamefont{Ravash}},
  \bibinfo{author}{\bibfnamefont{T.}~\bibnamefont{Feil}},
  \bibinfo{author}{\bibfnamefont{S.~D.} \bibnamefont{Danilov}},
  \bibinfo{author}{\bibfnamefont{J.}~\bibnamefont{Allerdings}},
  \bibinfo{author}{\bibfnamefont{D.}~\bibnamefont{Weiss}},
  \bibinfo{author}{\bibfnamefont{D.}~\bibnamefont{Schuh}},
  \bibinfo{author}{\bibfnamefont{W.}~\bibnamefont{Wegscheider}},
  \bibnamefont{and} \bibinfo{author}{\bibfnamefont{S.~D.}
  \bibnamefont{Ganichev}}, \bibinfo{journal}{Phys.\ Rev.\ Lett.}
  \textbf{\bibinfo{volume}{103}}, \bibinfo{pages}{090603}
  (\bibinfo{year}{2009}).

\bibitem[{\citenamefont{Grifoni et~al.}(2002)\citenamefont{Grifoni, Ferreira,
  Peguiron, and Majer}}]{Grifoni}
\bibinfo{author}{\bibfnamefont{M.}~\bibnamefont{Grifoni}},
  \bibinfo{author}{\bibfnamefont{M.~S.} \bibnamefont{Ferreira}},
  \bibinfo{author}{\bibfnamefont{J.}~\bibnamefont{Peguiron}}, \bibnamefont{and}
  \bibinfo{author}{\bibfnamefont{J.~B.} \bibnamefont{Majer}},
  \bibinfo{journal}{Phys.\ Rev.\ Lett.} \textbf{\bibinfo{volume}{89}},
  \bibinfo{pages}{146801} (\bibinfo{year}{2002}).

\bibitem[{\citenamefont{Goychuk and H{\"a}nggi}(1998)}]{Goychuk}
\bibinfo{author}{\bibfnamefont{I.}~\bibnamefont{Goychuk}} \bibnamefont{and}
  \bibinfo{author}{\bibfnamefont{P.}~\bibnamefont{H{\"a}nggi}},
  \bibinfo{journal}{EPL} \textbf{\bibinfo{volume}{43}}, \bibinfo{pages}{503}
  (\bibinfo{year}{1998}).

\bibitem[{\citenamefont{Ponomarev et~al.}(2009)\citenamefont{Ponomarev,
  Denisov, and H{\"a}nggi}}]{Ponomarev}
\bibinfo{author}{\bibfnamefont{A.~V.} \bibnamefont{Ponomarev}},
  \bibinfo{author}{\bibfnamefont{S.}~\bibnamefont{Denisov}}, \bibnamefont{and}
  \bibinfo{author}{\bibfnamefont{P.}~\bibnamefont{H{\"a}nggi}},
  \bibinfo{journal}{Phys.\ Rev.\ Lett.} \textbf{\bibinfo{volume}{102}},
  \bibinfo{pages}{230601} (\bibinfo{year}{2009}).

\bibitem[{\citenamefont{Rashba}(1960)}]{Rashba}
\bibinfo{author}{\bibfnamefont{E.}~\bibnamefont{Rashba}},
  \bibinfo{journal}{Fiz.\ Tverd.\ Tela (Leningrad)}
  \textbf{\bibinfo{volume}{2}}, \bibinfo{pages}{1224} (\bibinfo{year}{1960}).

\bibitem[{\citenamefont{Dresselhaus}(1955)}]{Dresselhaus}
\bibinfo{author}{\bibfnamefont{G.}~\bibnamefont{Dresselhaus}},
  \bibinfo{journal}{Phys.\ Rev.} \textbf{\bibinfo{volume}{100}},
  \bibinfo{pages}{580} (\bibinfo{year}{1955}).

\bibitem[{\citenamefont{Scheid et~al.}(2007)\citenamefont{Scheid, Pfund,
  Bercioux, and Richter}}]{Scheid}
\bibinfo{author}{\bibfnamefont{M.}~\bibnamefont{Scheid}},
  \bibinfo{author}{\bibfnamefont{A.}~\bibnamefont{Pfund}},
  \bibinfo{author}{\bibfnamefont{D.}~\bibnamefont{Bercioux}}, \bibnamefont{and}
  \bibinfo{author}{\bibfnamefont{K.}~\bibnamefont{Richter}},
  \bibinfo{journal}{Phys.\ Rev.\ B} \textbf{\bibinfo{volume}{76}},
  \bibinfo{pages}{195303} (\bibinfo{year}{2007}).

\bibitem[{\citenamefont{Smirnov
  et~al.}(2008{\natexlab{a}})\citenamefont{Smirnov, Bercioux, Grifoni, and
  Richter}}]{Smirnov}
\bibinfo{author}{\bibfnamefont{S.}~\bibnamefont{Smirnov}},
  \bibinfo{author}{\bibfnamefont{D.}~\bibnamefont{Bercioux}},
  \bibinfo{author}{\bibfnamefont{M.}~\bibnamefont{Grifoni}}, \bibnamefont{and}
  \bibinfo{author}{\bibfnamefont{K.}~\bibnamefont{Richter}},
  \bibinfo{journal}{Phys.\ Rev.\ Lett.} \textbf{\bibinfo{volume}{100}},
  \bibinfo{pages}{230601} (\bibinfo{year}{2008}{\natexlab{a}}).

\bibitem[{\citenamefont{Smirnov
  et~al.}(2008{\natexlab{b}})\citenamefont{Smirnov, Bercioux, Grifoni, and
  Richter}}]{Smirnov_1}
\bibinfo{author}{\bibfnamefont{S.}~\bibnamefont{Smirnov}},
  \bibinfo{author}{\bibfnamefont{D.}~\bibnamefont{Bercioux}},
  \bibinfo{author}{\bibfnamefont{M.}~\bibnamefont{Grifoni}}, \bibnamefont{and}
  \bibinfo{author}{\bibfnamefont{K.}~\bibnamefont{Richter}},
  \bibinfo{journal}{Phys.\ Rev.\ B} \textbf{\bibinfo{volume}{78}},
  \bibinfo{pages}{245323} (\bibinfo{year}{2008}{\natexlab{b}}).

\bibitem[{\citenamefont{Caldeira and Leggett}(1981)}]{Caldeira}
\bibinfo{author}{\bibfnamefont{A.~O.} \bibnamefont{Caldeira}} \bibnamefont{and}
  \bibinfo{author}{\bibfnamefont{A.~J.} \bibnamefont{Leggett}},
  \bibinfo{journal}{Phys.\ Rev.\ Lett.} \textbf{\bibinfo{volume}{46}},
  \bibinfo{pages}{211} (\bibinfo{year}{1981}).

\bibitem[{\citenamefont{Weiss}(2008)}]{Weiss}
\bibinfo{author}{\bibfnamefont{U.}~\bibnamefont{Weiss}},
  \emph{\bibinfo{title}{Quantum Dissipative Systems}}
  (\bibinfo{publisher}{World Scientific, Singapore}, \bibinfo{year}{2008}),
  \bibinfo{edition}{3rd} ed.

\end{thebibliography}
\end{document}